\documentclass[preprint,12pt]{elsarticle}

\usepackage{graphics,amssymb}

\journal{Nuclear Instruments and Methods A}

\def\megaohm{M$\Omega~$}
\def\ms{$\mu s~$}
\def\hplus{$\rm H^{+}~$}
\def\htwoplus{$\rm H_{2}^{+}~$}
\def\heplus{$\rm He^{+}~$}

\def\ntwoplus{$\rm N_{2}^{+}~$}
\def\oplus{$\rm O^{+}~$}
\def\otwoplus{$\rm O_{2}^{+}~$}
\def\chfourplus{$\rm CH_{4}^{+}~$}
\def\arplus{$\rm Ar^{+}~$}
\def\cotwoplus{$\rm CO_{2}^{+}~$}
\def\first{$\rm 1^{st}~$}

\begin{document}

\begin{frontmatter}



\title{Time and Amplitude of Afterpulse Measured with a Large Size Photomultiplier Tube.}


\author[sju]{K.~J. Ma}
\author[sju]{W.~G. Kang}
\author[pnu]{J.~K. Ahn}
\author[snu]{S. Choi}
\author[skku]{Y. Choi}
\author[sju]{M.~J. Hwang}
\author[cnu]{J.~S. Jang}
\author[sju]{E.~J. Jeon}
\author[cnu]{K.~K. Joo}
\author[cbu]{H.~S. Kim}
\author[cnu]{J.~Y. Kim}
\author[snu]{S.~B. Kim}
\author[cnu]{S.~H. Kim}
\author[knu]{W. Kim}
\author[sju]{Y.~D. Kim \corref{myaddress}}
\author[snu]{J. Lee}
\author[cnu]{I.~T. Lim}
\author[postech]{Y.~D. Oh}
\author[dsu]{M.~Y. Pac}
\author[skku]{C.~W. Park}
\author[gnu]{I.~G. Park}
\author[sku]{K.~S. Park}
\author[knu]{S.~S. Stepanyan}
\author[sku]{I. Yu}

\cortext[myaddress]{Corresponding author. E-mail:ydkim@sejong.ac.kr}
\address[snu]{Department of Physics, Seoul National University, Seoul 151-742, Korea}
\address[sju]{Physics Department, Sejong University, Seoul 143-747, Korea}
\address[cnu]{Department of Physics, Chonnam National University, Kwangju, 500-757, Korea}
\address[cbu]{Department of Physics, Chonbuk National University, Jeonju, 561-756, Korea}
\address[dsu]{Department of Physics, Dongshin University, Naju, 520-714, Korea}
\address[skku]{Department of Physics, Sungkyunkwan University, Suwon, 440-746, Korea}
\address[knu]{Department of Physics, Kyungpook National University, Daegu, 702-701, Korea}
\address[gnu]{Department of Physics, Gyeongsang National University, Jinju, 660-701, Korea}
\address[pnu]{Department of Physics, Pusan National University, Busan, 609-735, Korea}
\address[sku]{School of General Education, Seokyeong University, Seoul, 136-704, Korea}
\address[postech]{Department of Physics, Pohang University of Science and Technology, Pohang, 790-784, Korea}

\begin{abstract}
We have studied the afterpulse of a hemispherical photomultiplier tube for an upcoming reactor neutrino experiment.
The timing, the amplitude, and the rate of the afterpulse for a 10 inch photomultiplier tube 
were measured with a 400 MHz FADC up to 16 \ms time window after the initial signal generated by
an LED light pulse.
The time and amplitude correlation of the 
afterpulse shows several distinctive groups. We describe the dependencies of the afterpulse on the
applied high voltage and the amplitude of the main light pulse. The present data could shed light upon
the general mechanism of the afterpulse.
\end{abstract}

\begin{keyword}
Afterpulse \sep Photomultiplier Tube \sep Ionization
\PACS 29.40.Mc \sep 85.60.Ha


\end{keyword}

\end{frontmatter}


\section{Introduction}
\label{introduction}
Photomultiplier Tubes (PMT) are used in many areas of basic science research and
in commercial instruments for detection of low intensity light. Especially, nuclear and particle physics experiments
have used PMTs to detect even a single photon. 
PMTs are used to measure scintillation lights from various scintillators and the energy deposition
in the scintillator can be measured by integrating the charge of photoelectrons generated in the
photocathode of the PMTs. Afterpulse which occurs some time after the initial photoelectron signal is 
one of the  main sources of undesired background noise signals. The mechanism of afterpulse has been studied ever
since PMTs were used in the scientific research and the technology to reduce the afterpulse was critical to the PMT industry \cite{morton67}.
One of the main mechanisms generating the afterpulse was found to be the ionization of the residual gases by the accelerated
photoelectrons occurring inside the PMT \cite{morton67, coates73a}. The positive ions produced by the ionization 
are accelerated in the electric field towards the 
photocathode and generate an afterpulse upon the impact on the photocathode. Another mechanism of the
afterpulse is the back-scattered electrons at the PMT dynodes returning to the \first dynode after traveling
for a while inside the PMT. 

For experiments to measure very rare events, such as direct dark matter searches
or low energy neutrino experiments, PMT's afterpulse can be a significant source of the background since the signals
in these experiments are typically small and the event rate is very low.
Up to now, the afterpulse has been studied mainly with a conventional electronics such as a gated analogue to digital
converter (ADC) or a multihit TDC. For afterpulses occurring at wide range of time, the timing of the afterpulses
was not studied in detail. In this report, we describe a detailed study on both the precise time and the amplitude
of the afterpulses measured with a 400 MHz flash ADC (FADC). The offline analysis of the wide range of the
waveform enables the detailed studies on the mechanism of the afterpulse.
 
\section{Experimental Setup}
\label{}

\begin{figure}[h]
\centering
\includegraphics[width=17.0cm]{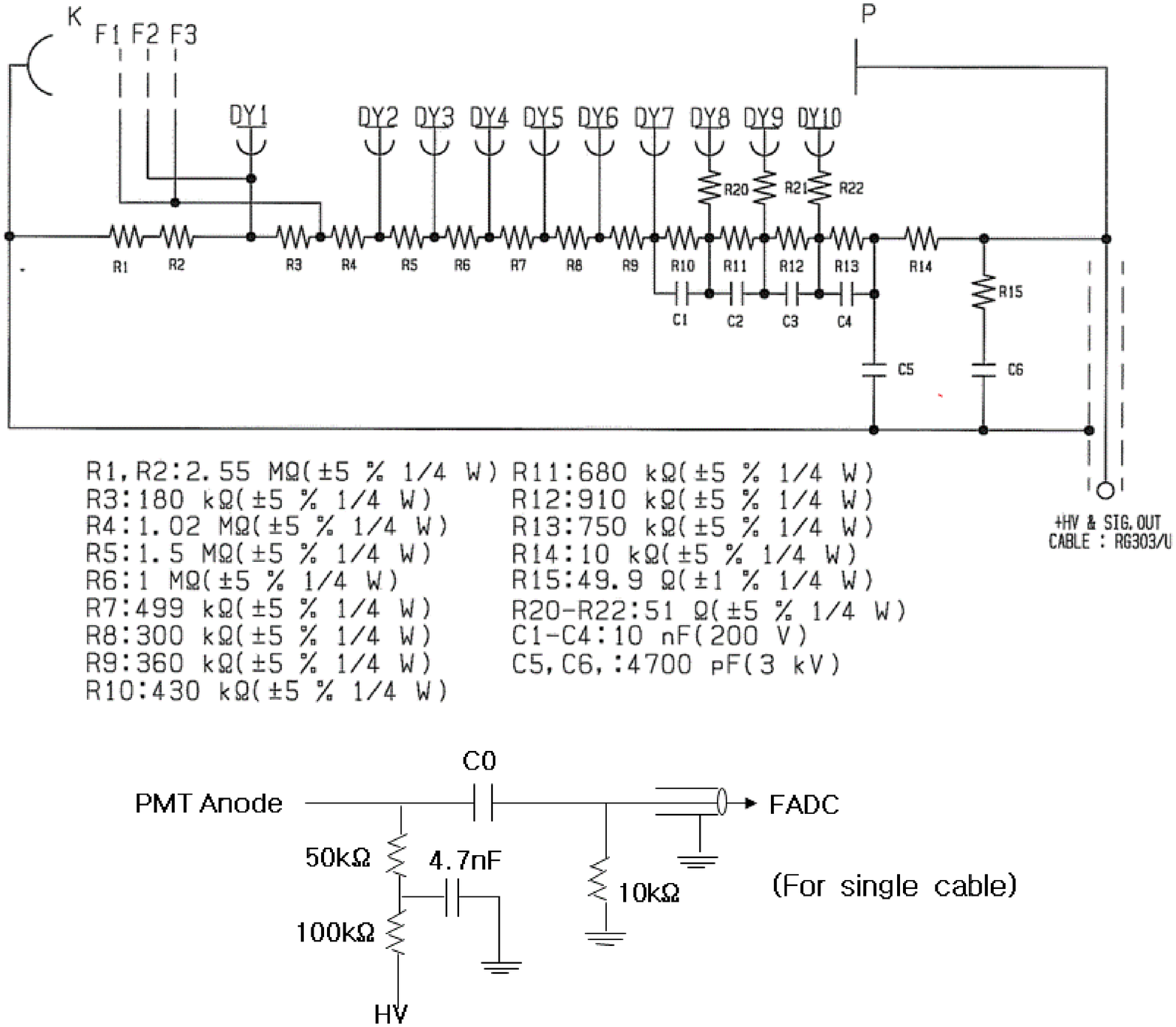}
\caption{The configuration of a voltage divider for R7081 PMT. A positive high voltage is applied and a single cable
is used for both high voltage and signal. Lower figure is the 
decoupling cicuit for the single cable.}
\label{basedrawing}
\end{figure}

For this measurement, a 400 MHz FADC module has been used to store the waveform of the 
PMT signal. 
Photons were generated by a light source made with a blue LED which was triggered by a short pulse 
from an electric pulse generator. 
The pulse width was 10 ns and the height of the pulse has been varied to control the light intensity from the LED.
We have tested a 10 inch hemispherical PMT(R7081 by Hamamatsu), which is being considered as photon detector
for a planned reactor neutrino oscillation experiment \cite{reno}.
The voltage divider of the PMT is shown in Figure \ref{basedrawing}. 
The total resistance of the base was 
12.7 \megaohm and the division ratio in the dynodes resistance chain was 
(16.8-0.6-3.4-5-3.33-1.67-1-1.2-1.5-2.2-3-2.4).
The voltage difference between the photocathode and the \first dynode is 622V for an applied voltage of 1550 V.

The FADC module was triggered by the pulse generator which provided light in the LED.
The duration of one event was 20.48 \ms (8192 bin and 2.5 ns per bin). 
To obtain the single photoelectron spectrum, the light intensity was adjusted so that there is about one signal 
out of 10 LED trigger events. The starting time of the LED light signal was about 4 \ms in 20.48 \ms length for one event.
Since the pulse width was only 10 ns, the main pulse occurred in the window of 50 ns, and the charge from the main pulse 
was obtained by integrating the charge centered at the maximum signal height of the pulse with the integration time window
of 40 ns. The afterpulse was searched in the time window of 300 ns - 15 \ms after the initial pulse. First, the local maximum
was searched and the 40 ns charge integration was performed for one afterpulse. 

\section{Results}

\subsection{Single photoelectron spectra}
\begin{figure}[h]
\centering
\includegraphics[width=10.0cm]{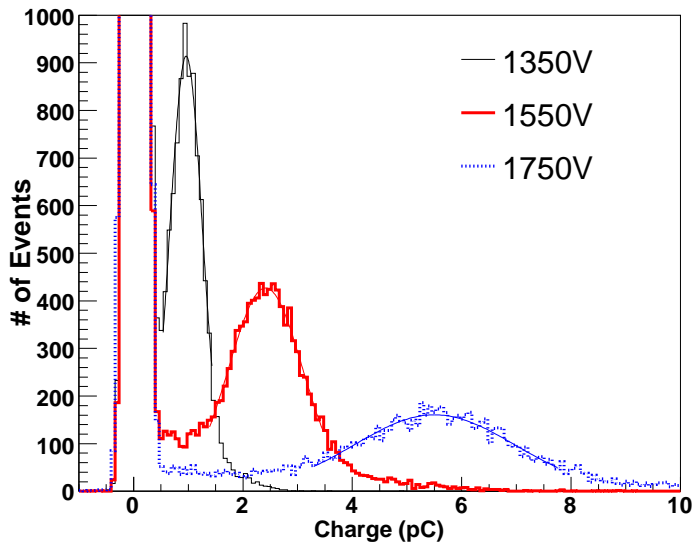}
\caption{Single photoelectron spectra for HV of 1350, 1550, and 1750. The thin lines on 
top of the spectra are the single gaussian fitting of the spectra.}
\label{spe}
\end{figure}

To study the mechanism of the various groups of afterpulses, different high voltage (HV) has been applied on the
PMT. The applied HV was varied from 1350 V to 1750 V. For each HV supply, a single photoelectron (SPE) spectrum was
obtained and the peak position gives the mean charge of single photoelectron. Then the number of photoelectrons (NPE) in 
the main pulse and each afterpulse
was determined by dividing the charge of each pulse with the SPE charge.
The SPE spectra for the three different high voltages are shown in Figure \ref{spe}.
The peak to valley ratio was 2.5 at 1350 V and more than 4 at 1750 V.

Table \ref{hvtable} lists the high voltages applied,
the SPE charge, and the mean number of photoelectrons in the main LED light pulse. A slight increase has been observed
in the number of photoelectrons for higher high voltages. This may be due to the higher electron collection efficiency.

\begin{table}
\begin{center}
\begin{tabular}{cccccc}\hline
High Voltage &   SPE charge     &  Gain              & Amplitude of LED pulse \\
 (V)         &  (pC)            &  ($10^7$)  & (NPE) \\ \hline
1350     &  0.962 $\pm$ 0.005   & 0.601 &   14.00\\
1400     &  1.243 $\pm$ 0.005   & 0.777 &   14.17\\
1450     &  1.549 $\pm$ 0.006   & 0.968 &   14.74\\
1500     &  1.964 $\pm$ 0.032   & 1.034 &   14.76\\
1550     &  2.414 $\pm$ 0.009   & 1.509 &   15.42\\
1600     &  2.964 $\pm$ 0.011   & 1.853 &   15.48\\
1650     &  3.639 $\pm$ 0.015   & 2.274 &   15.52\\
1700     &  4.456 $\pm$ 0.019   & 2.785 &   15.67\\
1750     &  5.503 $\pm$ 0.023   & 3.439 &   15.53\\ \hline
\end{tabular}
\end{center}
\caption{The charge of single photoelectron and the number of photoelectrons in the main LED light pulse for 
various high voltage applied to the PMT.}
\label{hvtable}
\end{table}

\subsection{Correlation of time and amplitude of afterpulses}
\label{timingsection}
\begin{figure}[h]
\centering
\includegraphics[width=14.0cm]{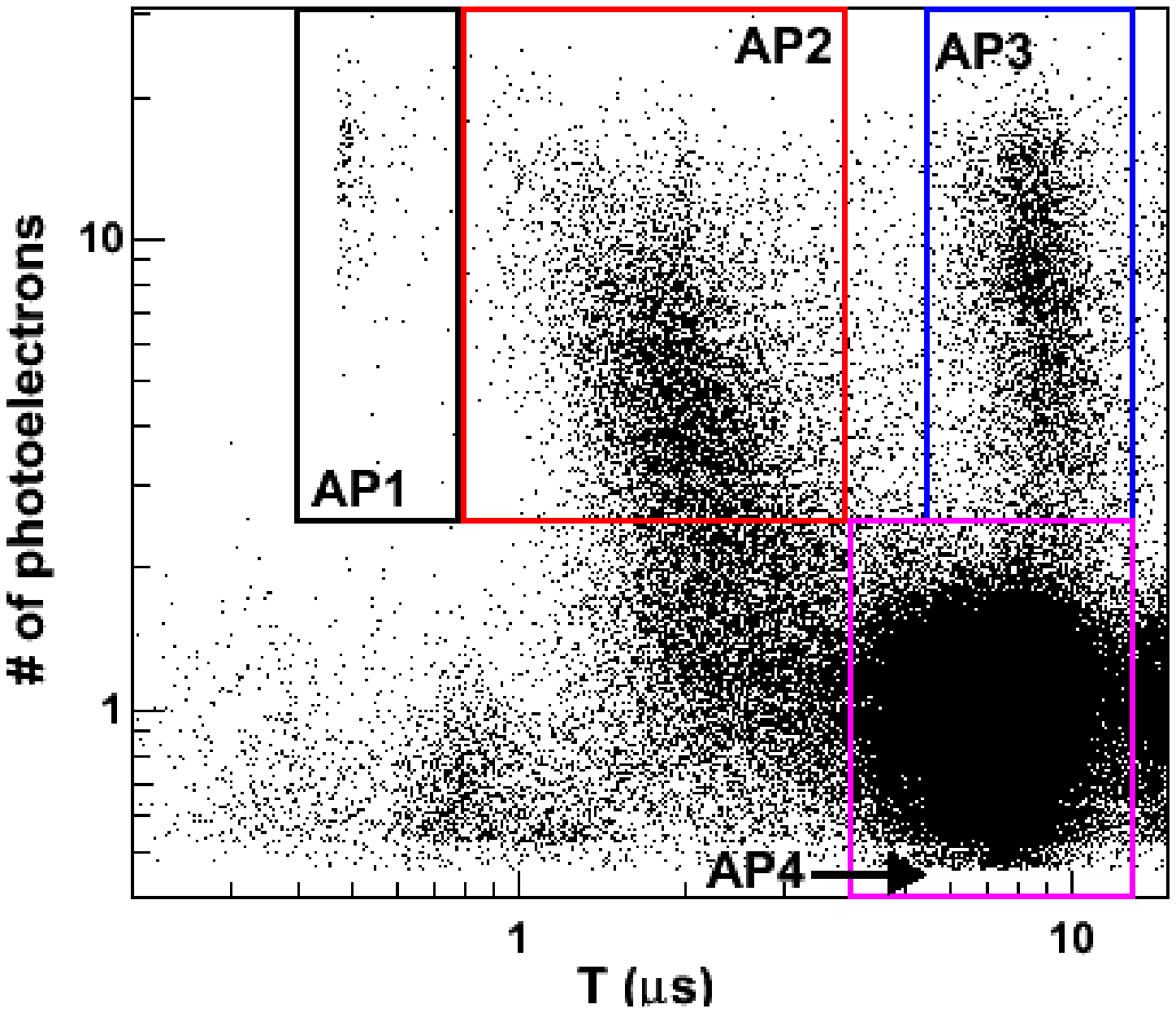}
\caption{The two dimensional scatter plot of the afterpulses. The x-axis is the occurring time in $\mu s$,
and the y-axis is the amplitude in NPE of the afterpulses. The boxes indicate the region for four groups in Table \ref{grouptable}. }
\label{npe_t}
\end{figure}

Figure \ref{npe_t} shows a scatter plot of NPE versus the arrival time of each afterpulse. 
The arrival time of the afterpulse is the time difference between the main pulse and each afterpulse.
There are
several groups which have distinct characteristics in time and amplitude. 
The afterpulse occurring at about 0.5 \ms has a large NPE value and a sharp time distribution. Afterpulse
between 1 \ms and 4 \ms has a wide distribution in amplitude from a single photoelectron upto tens
of photoelectrons. Afterpulse at around 8.5 \ms also has a wide distribution of NPE with a broad time
distribution. The most frequent afterpulses are single photoelectrons occuring at around 7.5 \ms.
Though the results described in this report is obtained with a R7081 PMT, we observed qualitatively analogous
results with another R7081 PMT. 

To investigate the afterpulses in detail, we have categorized the afterpulses into
four groups in the time-amplitude plot. Table \ref{grouptable} shows the ranges in the time and amplitude distribution
of the four groups. Though there are other afterpulses outside of the four groups, we will concentrate on the
four groups to study in detail. The afterpulses belonging to the four groups are the majority of the afterpulses.

\begin{table}
\begin{center}
\begin{tabular}{cccccc}\hline
Group  &  Time Range  & Amplitude Range  & $<t>$ for 1550 V \\
       &  (\ms)       & NPE              &    (\ms)  \\ \hline
AP1   &  0.4-0.78 & 2.5-30.5  & 0.49 \\ 
AP2   &  0.8-3.9  & 2.5-30.5  & 2.06 \\
AP3   &  5.5-13.0 & 2.5-30.5  & 8.73 \\
AP4   &  4.0-13.0 & 0.4-2.5   & 7.48 \\ \hline
\end{tabular}
\end{center}
\caption{The four afterpulse groups separated in time and amplitude plot.}
\label{grouptable}
\end{table}

\begin{figure}[h]
\centering
\includegraphics[width=14.0cm]{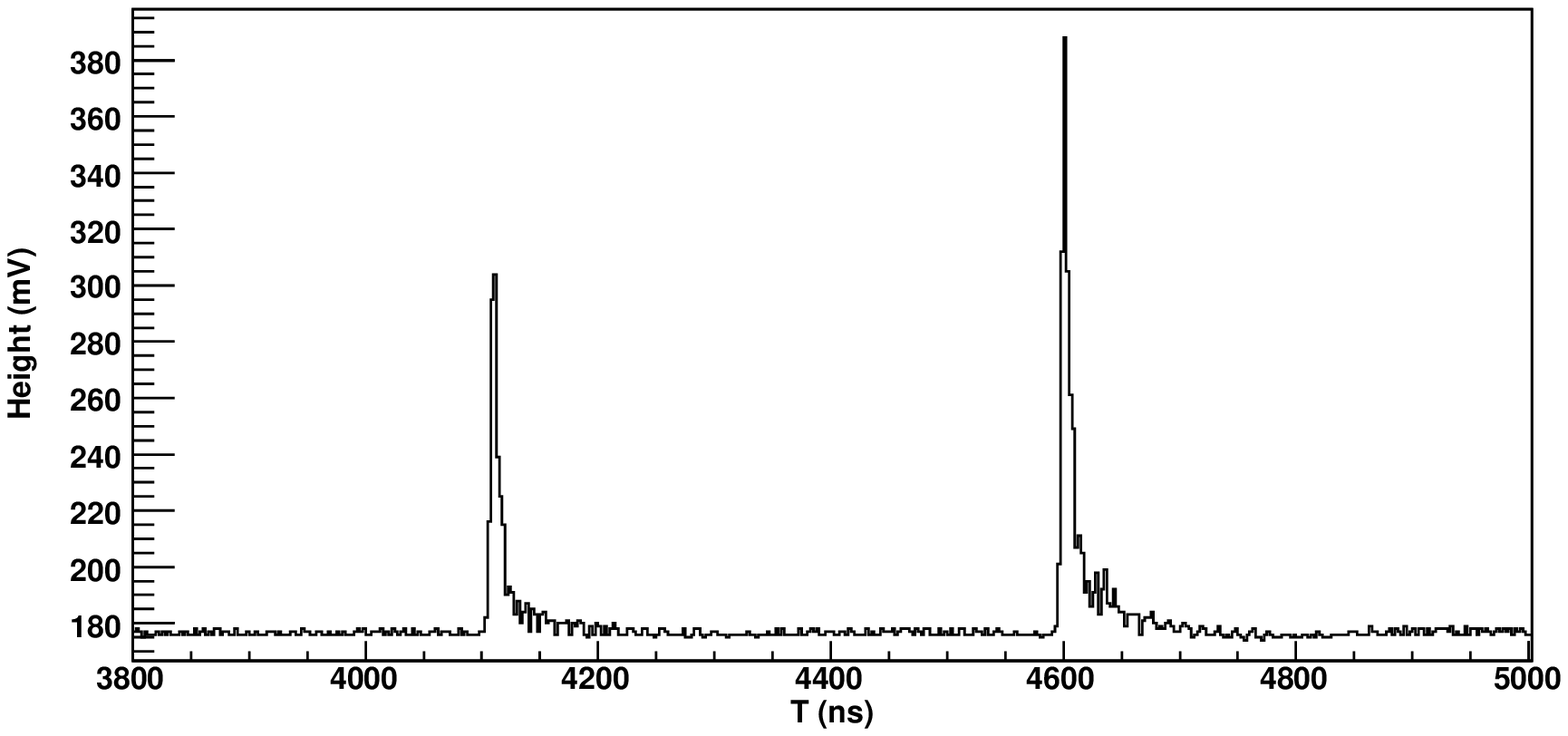}
\includegraphics[width=14.0cm]{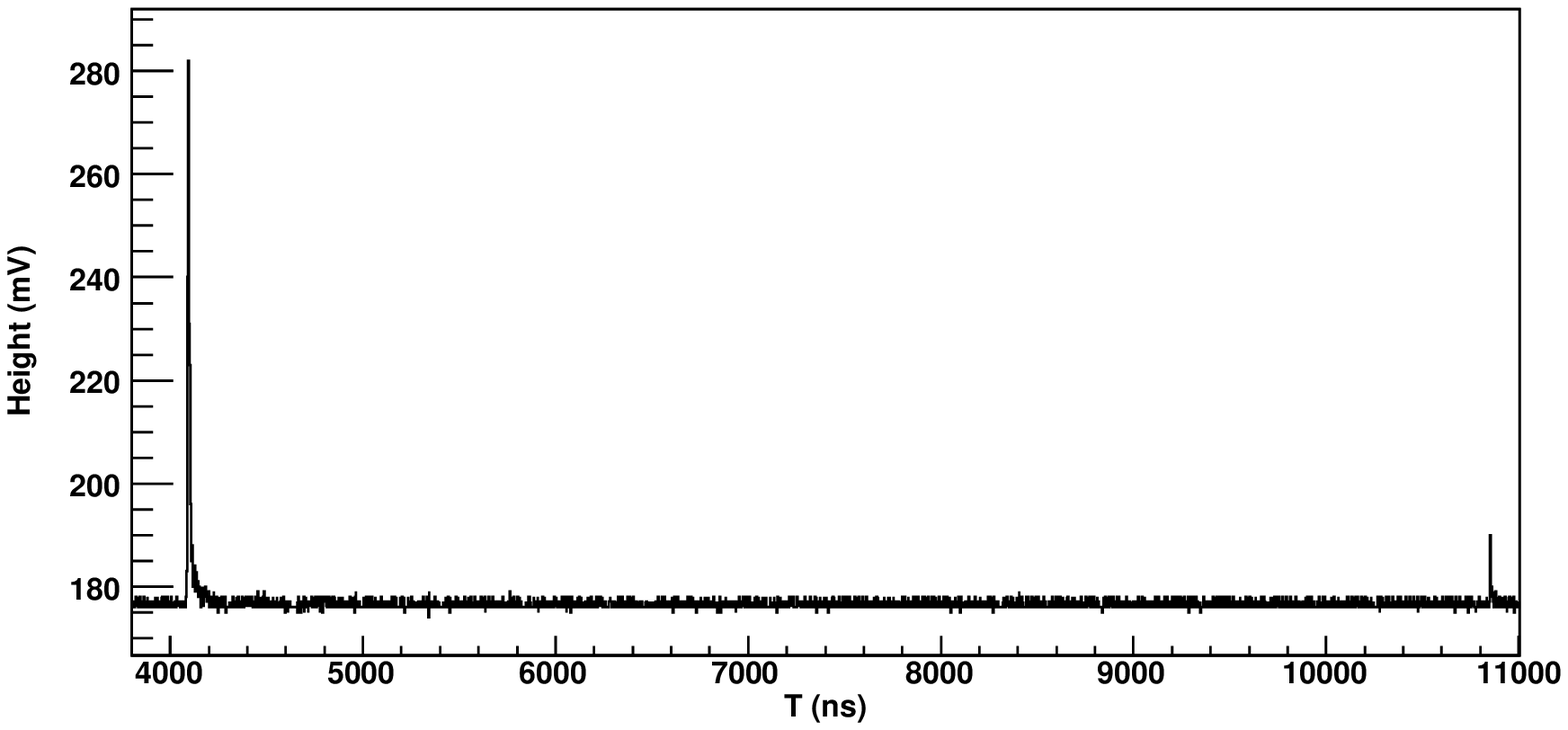}
\caption{ Examples of waveform for events with afterpulse. Upper figure is a waveform of an event with a large amplitude 
afterpulse of AP1 group(11.5 NPE of main pulse and 19.7 NPE of afterpulse). Lower figure is a waveform of an event 
with a SPE afterpulse of AP4 group.}
\label{wf}
\end{figure}

Two examples of the FADC waveform data are shown in Figure \ref{wf}. The top frame shows a waveform of an event
which has a large amplitude afterpulse of AP1 group(See Table \ref{grouptable} for the group specification). 
In this event the afterpulse has a larger signal than the main pulse.
The bottom frame shows a waveform of an event which has a SPE afterpulse of AP4 group.

\subsection{Afterpulse timing}
\label{}

The arrival time of ions produced inside the PMT depends on the electric potential distribution inside 
the PMT. Generally, the arrival time $\it t$ is \cite{coates73a,incandela87};
\begin{equation}
t= \int_{s_0}^{L} \frac{1}{v} ds = \sqrt{\frac{m}{2q}} \int_{s_0}^{L} [V(s_{0})-V(s)]^{-1/2} ds 
\end{equation}
Here, $\it m$ and $\it q$ are the mass and charge of the ion respectively, $s_{0}$ is the position of ionization,
$\it L$ is the position of photocathode, and $\it V(s)$ is the electric potential as a function of the position.   
If the electric potential distribution inside the PMT is linear as $V(s)=V_{0}(1-\frac{s}{L})$, then
the arriving time is 
\begin{equation}
t= \sqrt{\frac{2mL(L-s_{0})}{qV_{0}}}
\end{equation}
Here $V_{0}$ is the electric potential at the \first dynode and the electric potential at the photocathode is $\it V(L)=0$.

If the electric potential distribution inside the PMT is quadratic, $V(s)=V_{0}(1-\frac{s}{L})^{2}$, which is 
more realistic for a large size hemispherical PMT, the arrival time is given as
\begin{equation}
t= \sqrt{\frac{m}{2qV_{0}}}L \int_{s_0}^{L} \frac{1}{(L-s_{0})^{2}-(L-s)^{2}} ds = \frac{4}{\pi}\sqrt{\frac{2m}{qV_{0}}}L
\end{equation}
For this form of the electric potential distribution, the arrival time will be independent of the ionization position.

\begin{figure}[h]
\centering
\includegraphics[width=14.0cm]{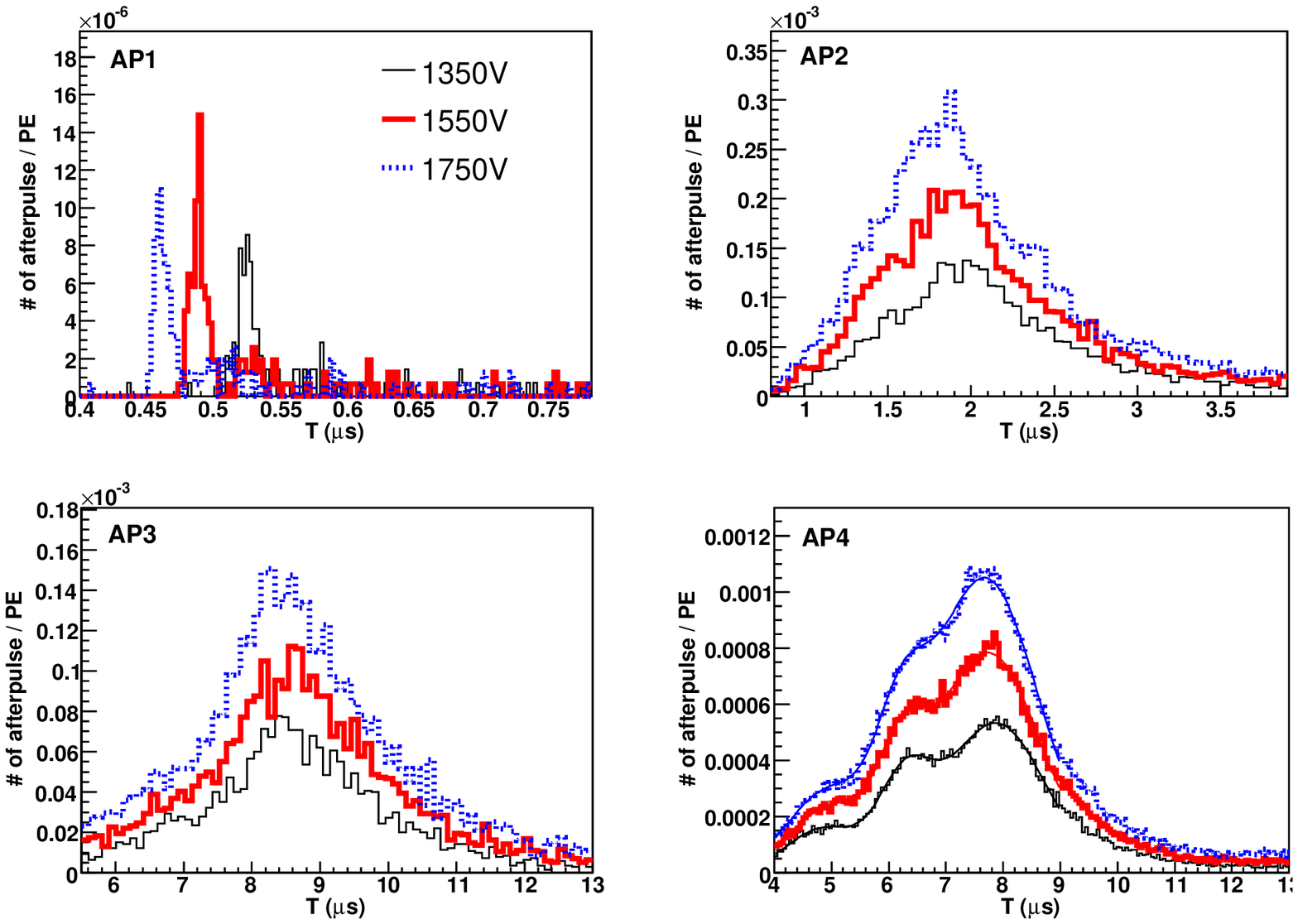}
\caption{The time distribution of the four afterpulse groups. For each group, the time distributions for HV of 
1350 V, 1550 V, and 1750 V are drawn. For AP4, the three gaussian fitting
 functions are drawn on top of the data spectra.}
\label{htcom}
\end{figure}

The arrival time of the afterpulse can be calculated more precisely with an actual map of the electric potential, but
the actual map is not available at the moment. Therefore the arrival time of the afterpulse in the data has been
compared with the simple model calculations.
The time distributions of the four afterpulse groups are shown in Figure \ref{htcom}. For each plot in the figure, 
three spectra taken
at different high voltages, 1350 V, 1550 V(green), and 1750 V(purple), are shown. 
The histograms are normalized by the number of LED trigger and the mean NPE in the LED pulse. Therefore,
the ${\it y}$ axis value is the probability to observe an afterpulse per single photoelectron at the photocathode per time bin
in each plot. As shown in 
the figure, the 
rates of the afterpulse of AP2, AP3, and AP4 increase as the applied high voltage increases (Section \ref{ratesection}). 
The arrival time of AP1 is especially sharp, centered at around 0.5 $\mu s$.

\begin{figure}[h]
\centering
\includegraphics[width=14.0cm]{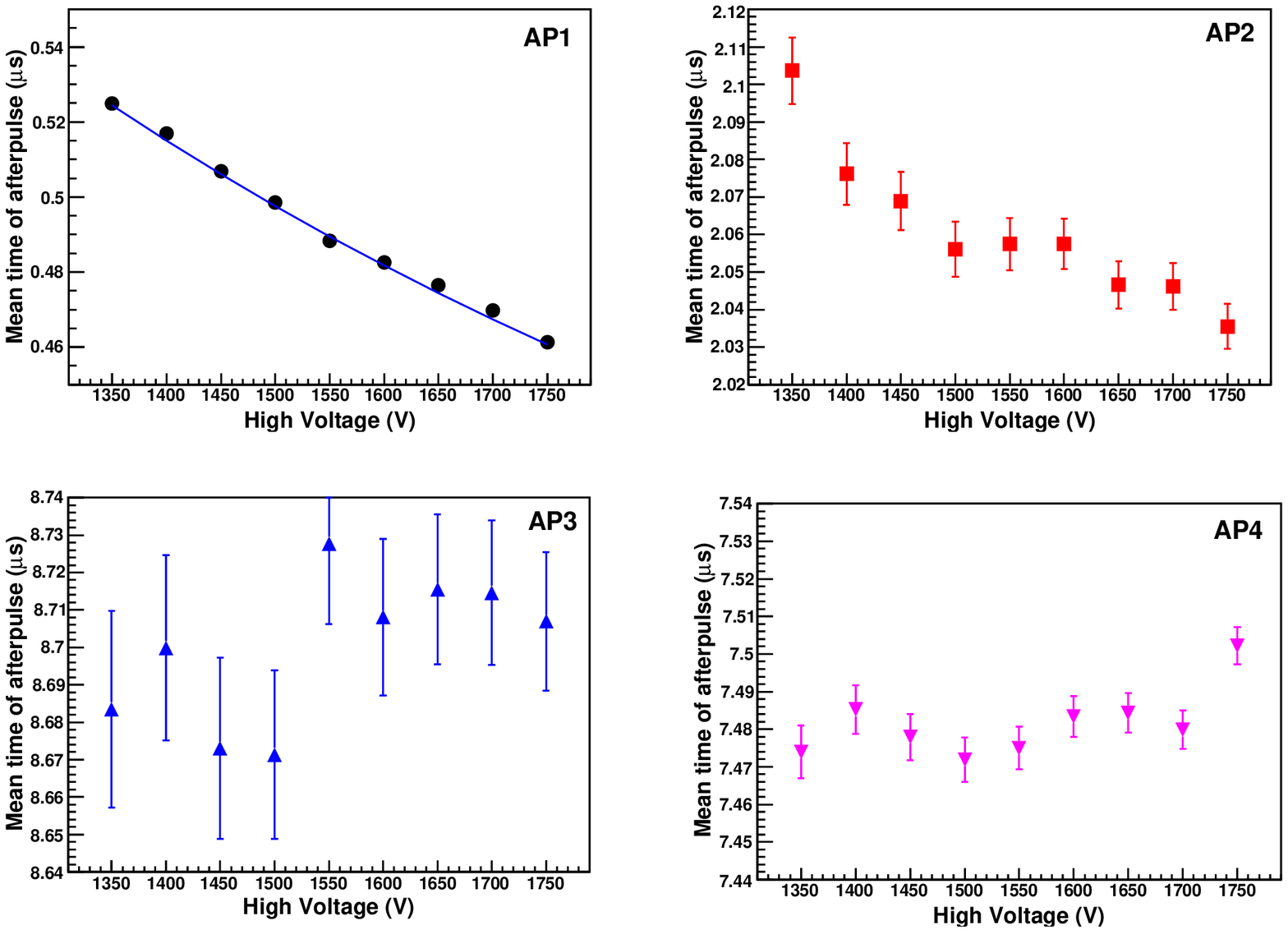}
\caption{The average time of all nine high voltage data for 
each afterpulse group. For all four figures the y-axis has the same range of 0.1 \ms. See the text for the line in AP1 data.}
\label{tmean_hv_com}
\end{figure}

The average arrival time of all nine high voltage data for 
each afterpulse group is shown in Figure \ref{tmean_hv_com}. The time distribution of AP1 group is fitted with a gaussian
function and the center value of the fitting function is drawn in Figure \ref{tmean_hv_com}. 
Afterpulses in AP1 group arrive earlier as the applied high voltage increases, which is a characteristic of
ion impact afterpulse. The distance between the center of photocathode and the \first dynode in R7081 PMT is known to be 16 cm, and 
the line in the AP1 plot is a theoretical calculation of equation (1) with $s_{0} $=11.5 cm when $\it L$=16 cm.
The line can also be a calculation of equation (3) as 
$t= \frac{4}{\pi}\sqrt{\frac{2m}{qV_{0}}}L \times 0.53$, where the factor 0.53 is necessary to explain the data.
Though the calculation reproduces the dependence of the AP1 timing very well, the $s_{0} $ value assuming the linear electric
potential distribution or the factor, 0.53, should not be seriously considered as an absolute value since the actual potential 
distribution would be different from the simple model.
However, the AP1 is the earliest afterpulse with 
high amplitude, and  we can ascribe this afterpulse to \hplus ion. The typical timing resolution of AP1 is about 6 ns in standard
deviation. This sharpness of AP1 afterpulse is not understood quantitatively yet, and will be discussed in Section \ref{discussion}.

Both of the AP2 and AP3 have large amplitude in continuous distribution. The arrival time of AP2 also gets faster
as the applied high voltage increases. The average arrival time of AP3 and AP4 is also drawn in Figure \ref{tmean_hv_com}.
These afterpulses have little dependence on the high voltage applied.
The arrival time distribution of AP4 afterpulse, which has an amplitude of single photoelectron, shows a structure 
with three bumps.
The arrival time spectra of AP4 was fitted with three gaussians and the fitting functions are drawn in Figure \ref{htcom} on top
of the data. This structure indicates that at least three different ions generate afterpulse with an amplitude of 
SPE. 
The fitting with three gaussian distributions reproduces well all the timing spectra with the different high voltages. 
Though the timing of three gaussian distribution has slight dependence on the high
voltage, the average arrival time of AP4 altogether has dependence on the high voltage less than 30ns. 

\subsection{Afterpulse amplitude}
\label{amplitudesection}
\begin{figure}[h]
\centering
\includegraphics[width=14.0cm]{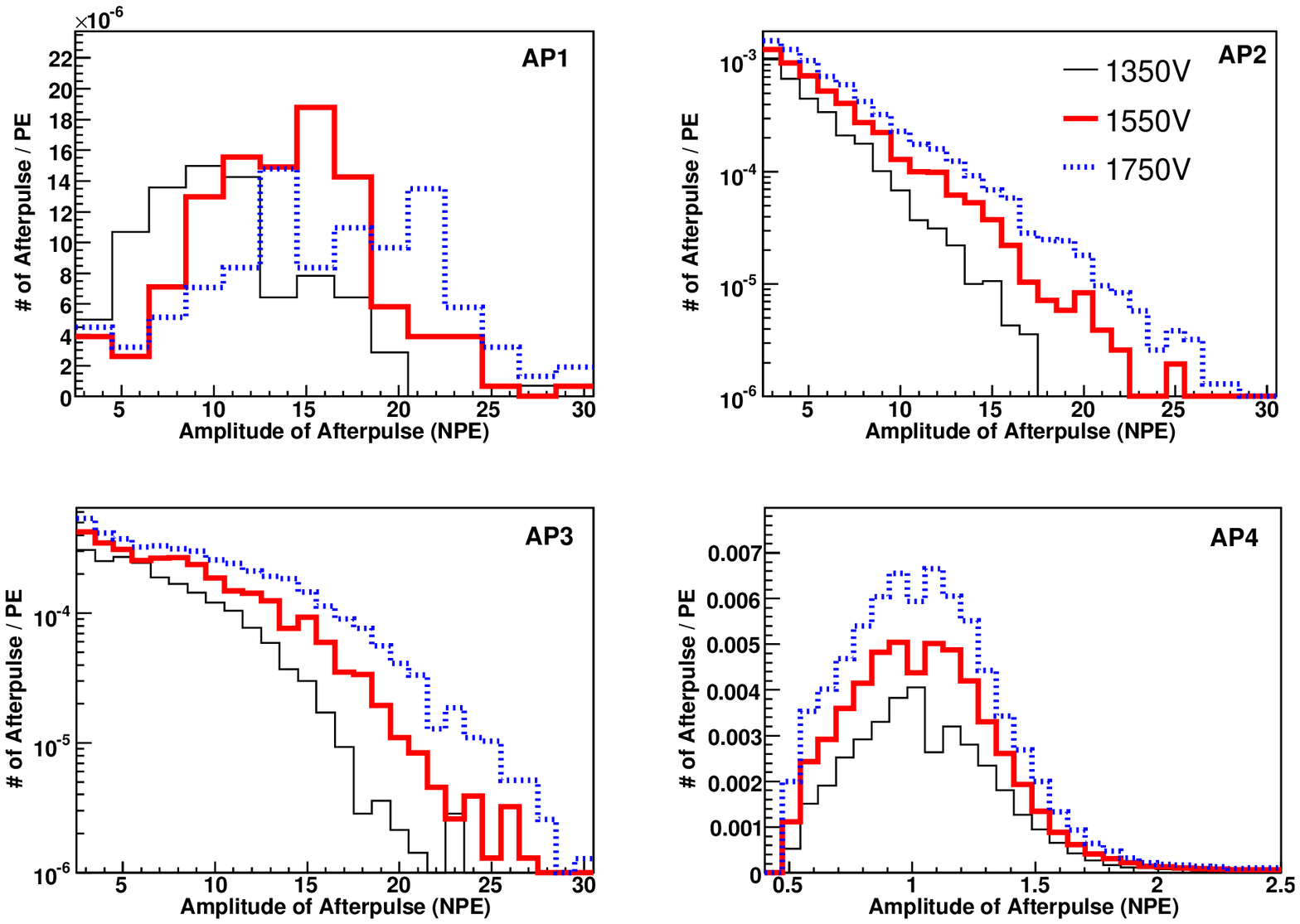}
\caption{The amplitude distributions of each afterpulse group in units of NPE. The y-axis value is
the probability of the afterpulse occrring with the amplitude per singlephoelectron in main pulse per each amplitude bin.
For each group, the data for HV of 
1350 V, 1550 V, and 1750 V are drawn.}
\label{hnpecom}
\end{figure}

\begin{figure}[h]
\centering
\includegraphics[width=10.0cm]{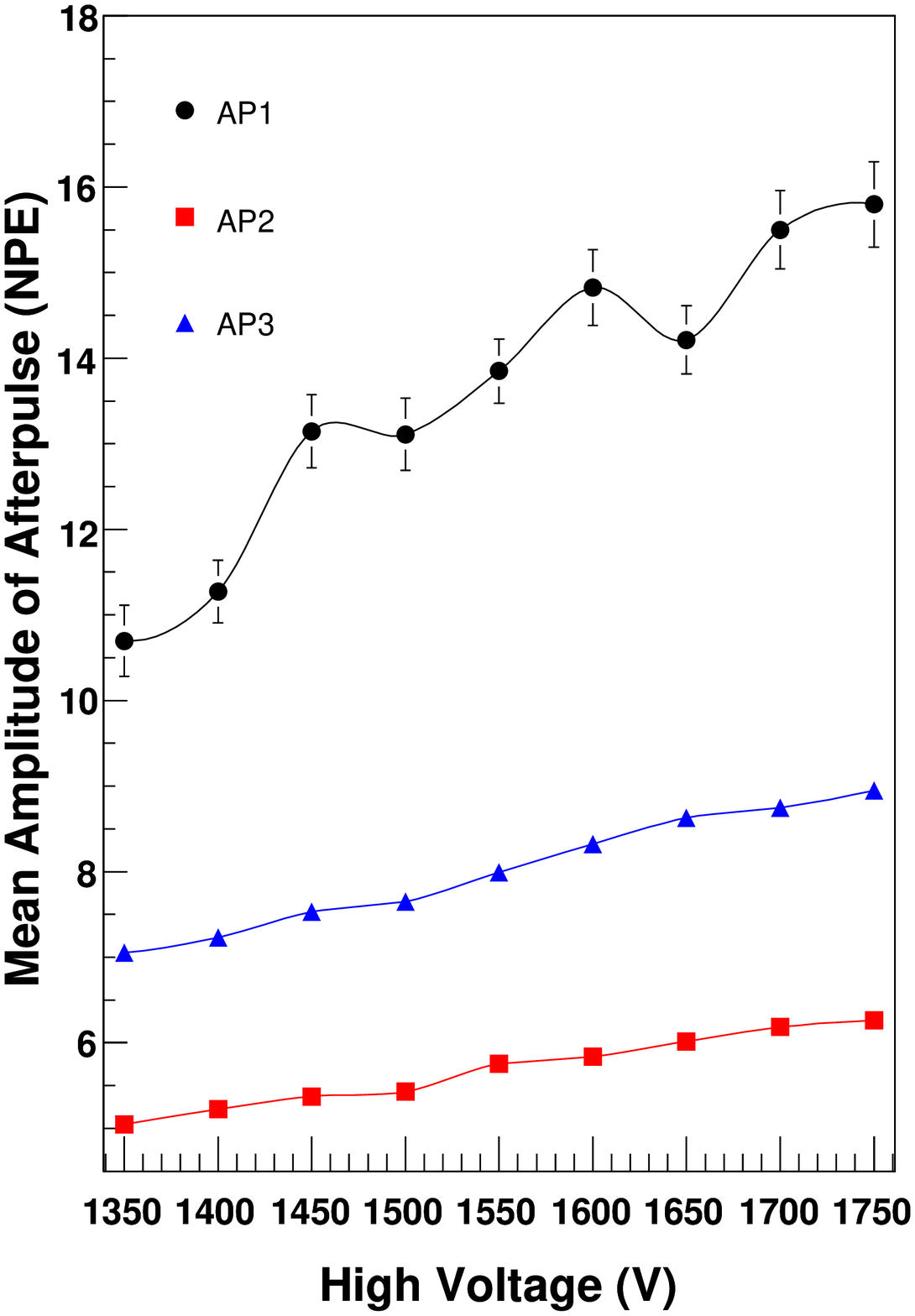}
\caption{The mean amplitude of the afterpulses as a function of applied high voltage. The amplitude of AP4 group is SPE.}
\label{npemean_hv}
\end{figure}

The amplitude distributions of each afterpulse group in units of NPE are shown in Figure \ref{hnpecom}. 
AP4 afterpulse is clearly a single photoelectron, and the amplitude distributions of AP2 and AP3 afterpulse are 
similar to Poisson distribution. The mean amplitude of the afterpulses is shown in Figure \ref{npemean_hv}
for all the values of the applied high voltage. The amplitude of AP1 afterpulse increaes by about 50 \% from 1350 V to 1750 V of
the applied voltage, and those of AP2 and AP3 increase by about 30 \%. 

\subsection{Afterpulse Rate}
\label{ratesection}
\begin{figure}[h]
\centering
\includegraphics[width=10.0cm]{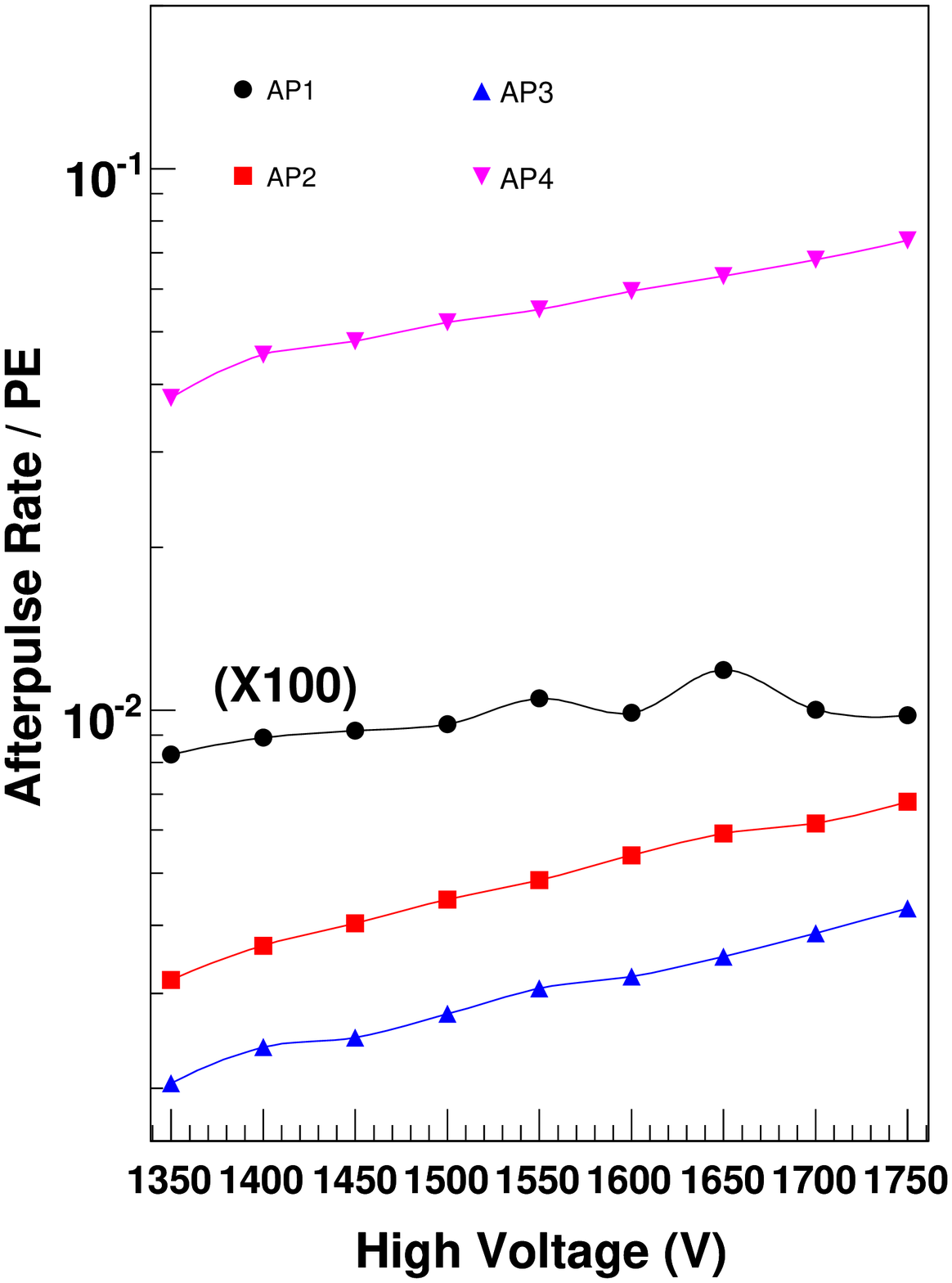}
\caption{Afterpulse rates for the four groups as a function of the applied high voltage. The AP1 data was multiplied by 100.}
\label{rate_hv}
\end{figure}

Most of the previous studies on the PMT afterpulse reported the rate of afterpulse as a 
function of time \cite{torre83,dornic06,ianni05}. They show that the afterpulse
peaks at around 6 \ms for 8 and 9 inch hemispherical PMTs \cite{dornic06, ianni05}.
Figure \ref{rate_hv} shows the afterpulse rate for the four groups as a function of the applied high voltage.
The rates are also normalized so that the rates are the probability per SPE in the main LED pulse, and they
are the integrated value of Figure \ref{htcom} or Figure \ref{hnpecom}. The rate of AP1 was multiplied by a factor
of 100 to make the curve more visible.
The rates of AP2, AP3, and AP4 increase as the applied high voltage increases. The rate of SPE afterpulse(AP4) is larger
than those of AP2 and AP3 by more than 10 times.

\subsection{Amplitude and Rate dependence on the Main Pulse Amplitude}
\begin{figure}[h]
\centering
\includegraphics[width=10.0cm]{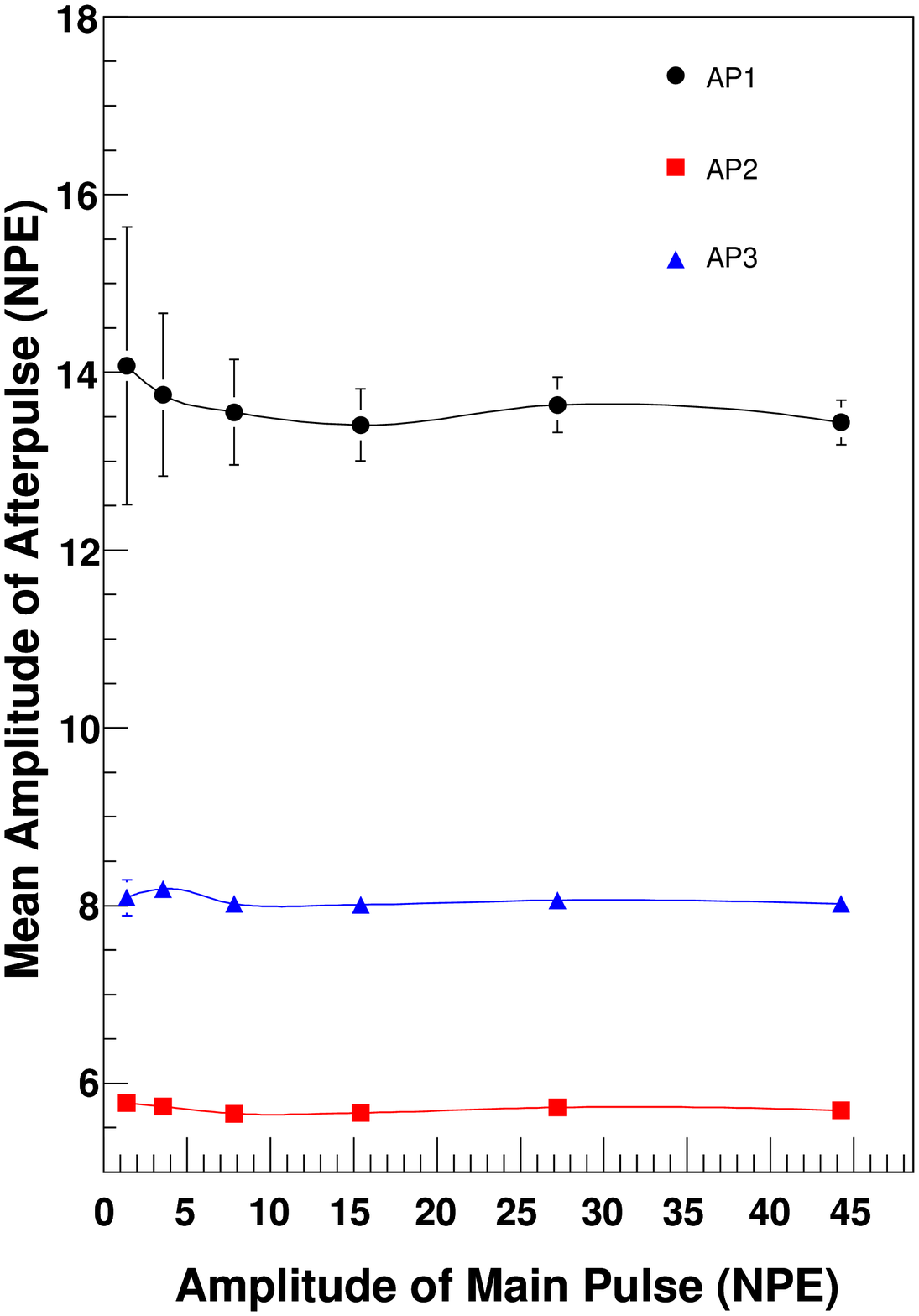}
\caption{Mean of afterpulse amplitudes as a function of the amplitude of the main pulse.
Both amplitudes are in the unit of NPE.}
\label{npemean_led}
\end{figure}

Since the ion impact afterpulse is due to the impact of accelerated ion toward the photocathode,
 the amplitude of inidividual afterpulse may depend on the
type of ion, but not on the amplitude of the main signal. Moreover, the rate
of afterpulse is expected to be proportional to the amplitude (or the number of photoelectrons) of the main signal.
We investigated the amplitude and rate dependence of individual afterpulse on the amplitude
of the main pulse by changing the light intensity of the LED. 
The LED driving pulse
height was increased in many steps up to the light level of 50 photoelectrons in the LED light pulse.
The high voltage was fixed at 1550 V which
gives a gain of $1.5 \times 10^{7}$. 
Figure \ref{npemean_led} shows the average number of photoelectrons of AP1, AP2, and AP3 afterpulse as a function of
the amplitude of the main pulse, which was varied from 1.5 up to 45. All the three afterpulse amplitudes are close 
to constant as expected. Though there are events which have more than one afterpulse, we can
separate the afterpulses thanks to the FADC.

\begin{figure}[h]
\centering
\includegraphics[width=10.0cm]{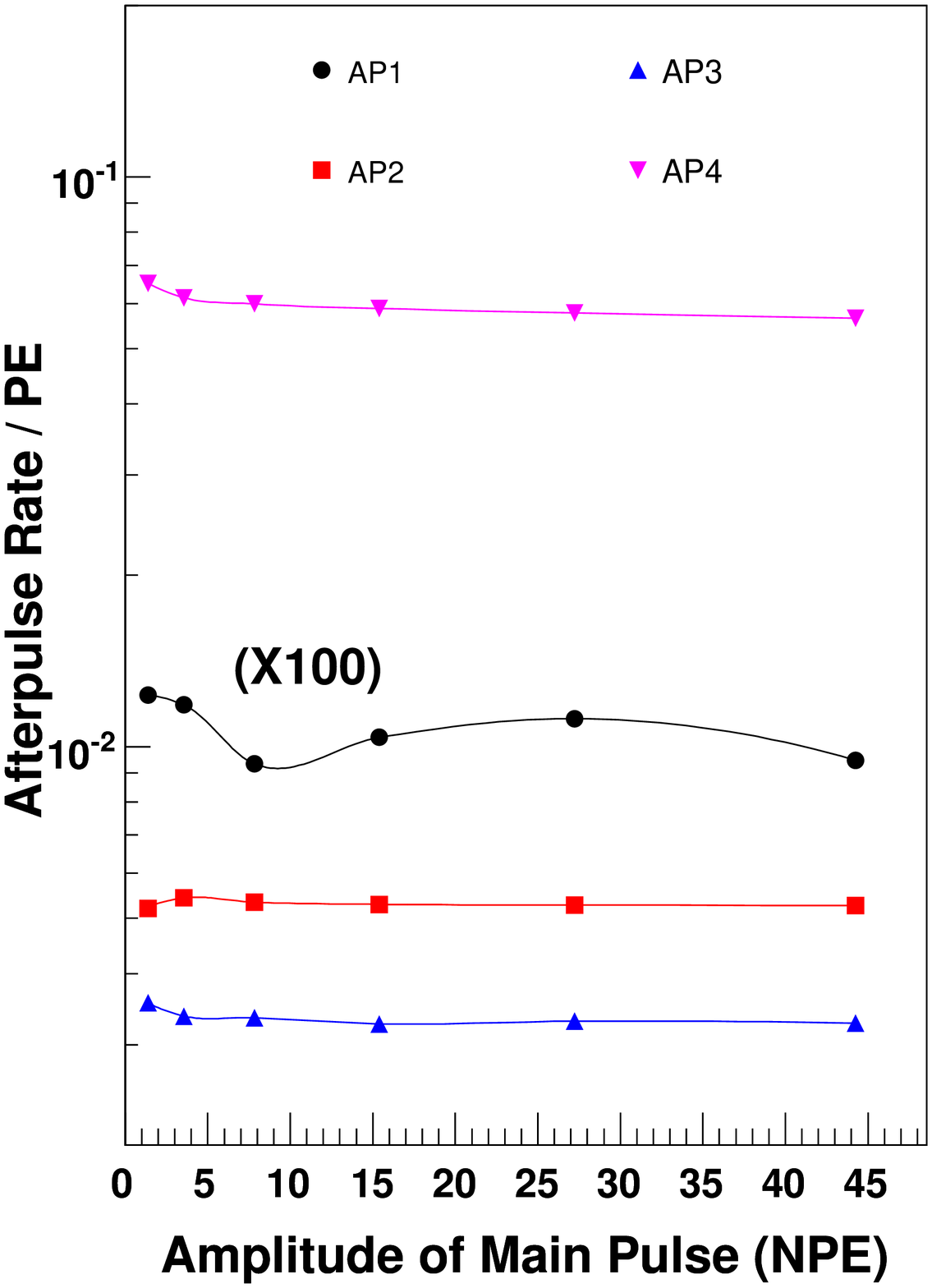}
\caption{Afterpulse rates for the four groups as a function of the amplitude of the main pulse. 
The AP1 data was multiplied by 100.}
\label{rate_led}
\end{figure}

The rates of afterpulse of AP1, AP2, and AP3 are less than 1\% for a single photoelectron in the main pulse.
Figure \ref{rate_led} shows the rate dependence on the amplitude of the main pulse. Basically, the normalized rates
(afterpulse rate per single photoelectron in main pulse) are independent of the amplitude of the main pulse. This result 
is also consistent with our expectation for ion impact afterpulse.
 
\section{Discussion}
\label{discussion}
The origin of the large amplitude afterpulse has been reported previously mainly as the ionized positive ion impact on the
photocathode \cite{morton67,coates73a, coates73b, torre83, akchurin07}. Morton et al. already reported that the amplitude of 
\htwoplus ion afterpulse is about four photoelectrons \cite{morton67} and Akchurin et al. recently reported that the amplitude
of three groups of afterpulse had the mean amplitude of 7 - 11 NPE.
Our results can be summarized as below.
\begin{itemize}
\item The amplitudes of the ion impact afterpulse increase as the applied HV, but do not depend on the amplitude of the main pulse.
\item The rates of the ion impact afterpulse increase as the applied HV, and increase linearly as the amplitude of main pulse.
\item The amplitude of the most frequent afterpulse is the same as single photoelectron, and their arrival time does not 
depend on the applied HV.
\end{itemize}
The first and the second results are consistent with \cite{morton67, akchurin07}.

The SPE afterpulse (AP4) should be due to the ion impact since they occur at around 7 \ms and we understand that
the backscattered electrons at the \first dynodes occur earlier than 100 ns for the PMT tested.
The electron transit time of the R7081 PMT is 63 ns.
Our data indicates that there are at least three different ions responsible for the AP4 afterpulse, and 
it will be very interesting to identify the ions which generate only a single electron at the impact.
Morton et al \cite{morton67} reports that nitrogen ion generates only single electron afterpulses unlike 
other ions.
Table \ref{timecompare} lists the measured mean time of each afterpulses at the applied high voltage of 1550 V
and the calculated timing of afterpulse with an equation (3) for various ions. 

\begin{table}
\begin{center}
\begin{tabular}{cccccc}\hline
Group  &  $<t>$ for 1550 V  & Ion (Calculated Time)\\
       &    (\ms)          &  (\ms)   \\ \hline
AP1   &  0.49 $\pm$ 0.001  & \hplus (0.92) \\ 
AP2   &  2.06 $\pm$ 0.01   & \heplus (1.85)\\
AP3   &  8.73 $\pm$ 0.02   & \cotwoplus (6.13)\\ 
AP4-1   &  5.19 $\pm$ 0.05 & \oplus(3.69), \chfourplus (3.69)\\ 
AP4-2   &  6.89 $\pm$ 0.02 & \ntwoplus (4.89), \otwoplus (5.22) \arplus (5.84) \\ 
AP4-3   &  7.70 $\pm$ 0.02 & \cotwoplus (6.13)\\ \hline
\end{tabular}
\end{center}
\caption{The comparison between timing of the afterpulses measured and timing calculated with Eq. (3) for various ions.}
\label{timecompare}
\end{table}

As discussed in Section \ref{timingsection}, the AP1 afterpulse can be ascribed to \hplus (proton) because
\hplus has the least $\frac{m}{q}$ among possible ions. The time distribution of the afterpulse is about 6 ns in sigma, 
and such a sharpness is difficult to understand. The arrival time of \hplus calculated with Eq.(3) is 0.91 \ms 
for the applied voltage of 1550 V which is far from the measured value of 0.49 $\mu s$.
If the ionization occurs between the photocathode and the \first dynode, the arrival time can not be so sharp since 
the actual potential distribution would be different from the assumed potential distribution of Eq. (3).
One possible explanation would be that AP1 afterpulse was due to the \hplus generated at the \first dynode with 
the adsorbed $\rm H_{2}O$ on the dynode.
We need a detail map of the electric potential inside the PMT to clarify the issue of the sharpness of the afterpulse.

In summary, we have studied the timing, amplitude, and the rate of PMT afterpulse for a wide range of the arrival times in detail.
The time and amplitude correlation of the 
afterpulse shows several distinctive groups. The afterpulse with single photoelectron amplitude 
was observed as separated from other
large amplitude afterpulse for the first time. The afterpulse potentially assigned as \hplus 
showed a time width less than 10 ns. The amplitude and normalized rate of ion impact afterpulse 
does not depend on the main pulse amplitude in the range of up to 45 photoelectrons.

This work is supported by the Korean Ministry of Education, Science and Technology, through the project 
of neutrino detector facility, by Korea Neutrino Research Center throuth National Research Foundation of Korea
Grant (2009-0083526), and by the Korea Research Foundation Grant funded by the Korean Government(MOEHRD) (KRF-2006-C00101).



\end{document}